\documentclass[12pt]{article}
\begin{document} 
\begin{titlepage}	
\begin{center}
{\Large {\bf Dynamic Scaling for First-order Phase Transtions$^\dagger$}} 
\end{center}
\vskip 1.5cm
\centerline{\bf Banu Ebru \"{O}zo\u{g}uz, Yi\u{g}it G\"{u}nd\"{u}\c{c} 
and Meral Ayd{\i}n}
\centerline{\bf Hacettepe University, Physics Department,}
\centerline{\bf 06532  Beytepe, Ankara, Turkey }
\vskip 0.5cm

\centerline{\normalsize {\bf Abstract} }

{\small The critical behaviour in short time dynamics for the $q=6$ and
$7$ state Potts models in two-dimensions is investigated. It is
shown that dynamic finite-size scaling exists for first-order phase
transitions.}

\vskip 0.4cm


\vskip 2.0cm

$\dagger${This project is partially supported by Turkish Scientific
and Technical Research Council (T\"{U}B\.{I}TAK) under the project
TBAG-1669.}
\end{titlepage}	

\pagebreak

\section{Introduction}

It has been shown that~\cite{Janssen:1989} for a dynamical
relaxation process in which a system, evolving according to a dynamics
of model A~\cite{Hohenberg:1977}, and is quenched from very high temperature
to the critical temperature, there emerges a universal dynamical
scaling behaviour already within the short-time regime.  This rather
unexpected scaling seems to exist, since spatial correlations and
correlation time diverge simultaneously as the system approaches long-
time regime at the critical temperature. For the short-time regime, 
the finite-size scaling form of the time evolution of a $k$-th moment of 
the magnetization is written as~\cite{Janssen:1989}

\begin{equation}\label{FSS_Mag}
M^{(k)} (t,\epsilon,m_0)\;=\;b^{- k \beta / \nu} \; M^{(k)} (t
/\tau(L),b^{1/\nu}\epsilon,b^{x_0} m_0)   .
\end{equation}

Here $b$ is the scale change, $L$ is the linear dimension of the
system, $\beta$, $\nu$ are the well known static critical exponents,
$\tau$ is the autocorrelation time, and ${\epsilon=(T-T_c)/T_c}$ is the
reduced temperature. Short-time dynamic behaviour also requires a new
independent critical exponent $x_0$ which is the scaling dimension of
the initial magnetization $m_0$. It has been shown numerically that
dynamic scaling exists~\cite{Zheng:1998} even at the very early stages of the
relaxation process.

$\;$

Rigorous formulation of the finite-size scaling for first-order phase
transitions~\cite{Borgs:1990,Borgs:1991} resulted in better
understanding of the dynamics of the first-order phase transitions. In
this formalism it has been shown that the phase transition is governed
by the surface tension between the ordered and disordered phases. The
system tunnels between these two metastable states and these
transitions are observed during simulation studies of the long-term
behaviour of the system. For finite systems, \-undergoing first-order
transitions, the autocorrelation time $\tau$ for the relaxation process has
been calculated~\cite{Binder:1982,Niel:1987} for cluster
algorithms~\cite{Wang:1987,Wolff:1989} and is given as

\begin{equation}\label{TAU_First_Order}
\tau \;=\; \displaystyle{{L^{d/2}}
\exp(\sigma_{od} \; L^{d-1})}   ,
\end{equation}

where $d$ is the dimensionality of the system.This form of $\tau$ can
be used to identify the order of the phase transition. 

$\;$

In a series of previous work~\cite{Aydin:1996,Yasar:1998}, behavioral
differences between first- and second-order phase transitions have
been studied. In these works, empirically distinct change in the time
evolution of the operator in initial stages of the simulation gave a
clear indication that first- and second-order phase transitions are
grouped into two different evolutionary processes. Since short-time
dynamic behaviour of second-order phase transitions are well
understood in terms of dynamic scaling formalism~\cite{Janssen:1989},
in analogy with second-order transition, scaling for first-order
transitions may be put into more rigorous footing. The success of the
finite size scaling arguments and explicit form of $\tau$ given in
Eq.(2) led us to study the existence of short-time dynamic scaling in
first-order phase transitions. In first-order phase transitions the
singularities are governed by the volume of the system. Hence in
first-order transitions the thermal and magnetic critical indices are
replaced by the dimension of the system.  Combining this information
with Eq.(2), we have formulated dynamic scaling form of various
operators in anology with Eq. (1). In this work, our aim is to show
that a system exhibiting first-order phase transition obeys well
defined dynamic finite-size scaling rules during quenching from
disordered state to the infinite lattice transition temperature.  We
have studied the short-time relaxation processes by using $q=6$ and
$7$ state Potts models, which are known to exhibit first-order phase
transition. In this model we studied the time evolutions of the order
parameter, the largest cluster and the Binder cumulant \cite{Binder:1981}.

$\;$

\section{Model and Method}

The Hamiltonian of the $q$-state Potts model~\cite{Potts:1952,Wu:1982} is
given as

\begin{equation}\label{PottsHamiltonian}
- \beta H \;=\; {\sum_{<ij>}} K \delta_{s_i, s_j}
\end{equation}

where the spin $s$ can take values $1,\dots,q$, $\beta = 1 /{k_B T}$
is the inverse temperature, $K= J/(k_B T)$, $\delta$ is the Kroneker
delta function, and sum is over all nearest-neighbour pairs on
two-dimensional lattice. In equilibrium the $q$-state Potts model is
exactly solvable.  The critical point locates at $K_c\;=\; \log(1 +
\sqrt{q})$. In principle, any type of dynamics can be given to the
system to study non-equilibrium time evolution process. In this work,
we used nonconserved dynamics of Model $A$~\cite{Hohenberg:1977}. In
order to study dynamic scaling in systems exhibiting first-order phase
transitions the following operators are considered:

\begin{enumerate}

\item Moments of the order parameter ($M$)

\begin{equation}
\label{magnetization}
M^{(k)}\;=\;\langle ({{q \; \rho^{\alpha}-1}\over {q-1}})^{k} \rangle
\end{equation}

$\rho^{\alpha}=N^{\alpha}/L^{d}$, $N^{\alpha}$ being the number
of spins with $s=\alpha$, $L$ the linear size and $d$ is the
dimensionality of the system.

\item Binder cumulant ($B$)

\begin{equation}
\label{binder}
B \;=\; 1 - {{ M^{(4)}}\over {3\, {M^{(2)}}^2}}
\end{equation}

\item Largest cluster ($C_m$)

\begin{equation}
\label{C_m}
C_{m} \;=\; {1\over {L^d}} \langle N_{C_m} \rangle
\end{equation}

\end{enumerate}

$N_{C_m}$ is the number of spins belonging to the largest cluster in
each configuration. Largest cluster gives the time evolution of the
average of the largest cluster found in each configuration. This
quantity scales like the susceptibility. Hence in a first-order phase
transition, it grows like volume.

$\;$

For the first-order phase transitions, since the static critical
exponents are replaced by the dimension of the system, rather than
calculating the static critical indices, one can test the validity of
the dynamic scaling assumption at the initial stages of the simulation
and obtain the surface tension as the result of the scaling. For the
computational simplicity, the initial magnetization $m_0$ is set to
zero. For second-order phase transitions, the finite-size behaviour of
the magnetization is given by Eq.(1). Here, $\beta/\nu\;=\;Y_H-d$.
Since $ Y_H $ and $Y_T$ are equal to the dimension of the system, for
first-order phase transitions, the order parameter ($M^{(1)}$), Binder
cumulant ($B$) and the largest cluster ($C_m$) scale according to

\begin{equation}
\label{scaling}
f_{L_1}(t/{\tau(L_1)},0,L_1)\;=\;f_{L_2}(t/{\tau(L_2)},0,L_2)
\end{equation}

where $\tau(L)$ is autocorrelation time of the lattice with the linear
size $L$. Application of this form to data can show scaling for
various size lattices. In the following section we have presented our
results.

\section{Results and Discussions}

 Following the considerations started in previous section, we have
 studied the two-dimensional $q=6$ and $7$ state Potts models evolving
 in time according to dynamics of model $A$~\cite{Hohenberg:1977}.
 Our main objective is to observe the dynamic scaling, hence we have
 prepared lattices with vanishing order parameter, avoiding the
 complications due to having an extra parameter $x_0$. This is achived
 for $q=6$ and $7$ state Potts models by choosing the lattice sizes as
 the integer multiples of $q$. Totally random initial configurations
 are quenched at the corresponding infinite lattice transition
 temperature. Simulations are performed on $6$ different lattice sizes
 by using Wolff cluster update algorithm. For each $q$ and $L$ the
 averages are taken over $10000$ different samples. Errors are
 calculated by dividing the samples into ten subsamples. As the
 lattice size grows, number of iterations for thermalization grows
 according to growing tunneling time (Eq. 2). For $q=7$ and larger
 lattices up to $30000$ iterations are necessary for
 thermalization. The chosen lattice sizes are $L=42,60,72,90,96,102$
 and $L=35,49,63,77,91,105$ for $q=6$ and $q=7$ respectively.

$\;$

The two-dimensional $q$-state Potts model is known to undergo
first-order phase transition for $q\; > \;4$~\cite{Baxter:1973}. Even
though the $q=7$ state Potts model exhibits strong first-order
behaviour, the correlation length is about $50$ lattice sites. Hence
for $q=7$~\cite{Buffenoir:1993}, the largest lattices are expected to
show good scaling behaviour without any need to correction to scaling
terms. For smaller lattices, however, one needs to consider the
correction to scaling according to the finite-size scaling theory for
first-order phase transitions. The general form of the corrections to
the scaling can be given as polynomial in ${1 \over {L^d}}$, which can
be written as

\begin{equation}
\left< A \right>_{L}\; = \; A_0 \;( { 1 \; + \;{ {A_1} \over {L^{d}} }
\; + \; { {A_2} \over {L^{2d}} }\; +\; \dots })   .
\end{equation}

This form indicates that all of the observables scale if one
calculates $A_0$ by fitting the correction to scaling
terms~\cite{Borgs:1990,Borgs:1991,Billoire:1995}. The correction to
scaling plays even more profound role for $q=6$ state Potts model
where the correlation length is larger than even the largest
lattice. The correction to scaling for each observable is obtained by
fitting the averages, taken over $10000$ iterations after the
thermalization, to Eq. (8) and the expansion coefficients $A_1,
A_2$,.. are calculated for $q=6$ and $7$.

$\;$

In Figure 1.a, the time evolution of the order parameter is plotted
for $q = 6$. As one can observe, for each lattice size, starting from
totally random configuration $m_0\,=\,0$, the order parameter evolves
to a plato. For large enough lattices, since $Y_H - d$ vanishes for
first-order phase transitions, one can expect the same long-term
behaviour for all different lattice sizes. In fact this is the case,
within the errorbars, for our largest two lattices. In order to see
scaling for smaller lattice sizes we have performed long runs, after
thermalization, and the correction to scaling terms (Eq. 8) are fitted
to the order parameter values. In figures 1.a and 1.b the raw data and
the scale form is presented for $q=6$ and $7$ respectively. Figures
1.c and 1.d show the scaled forms of the data in figures 1.a and 1.b
respectively. For $q=7$, the correction to scaling is almost
negligible for lattices larger than $L=65$.

$\;$

Similarly, for the averages of the maximum cluster, which is expected
to grow like the volume, we have observed that similar scaling
behaviour exists. Figures 2.a and 2.b are plots of the Monte Carlo data
for $q=6$ and $q=7$ respectively. Figures 2.c and 2.d are the scaled
form of the above mentioned data.

%$\;$

The last quantity that we have observed is the Binder cumulant.
Binder cumulant is a scaling function and also is a ratio of two
quantities of equal anomolous dimensionality. Hence, correction to
scaling terms are almost negligible even for very small lattices. 

\begin{table}
\begin{center}

\begin{tabular}{|c|c c c|c c c|}
\hline
           &q\,\, &  $=$   &6\,\,     & \,\, q  &   $=$    &7\\
\hline
   M      &0.0084& $\pm$  &    0.0011&    0.0185& $\pm$   &0.0015 \\
   B      &0.0076& $\pm$  &    0.0004&    0.015 & $\pm$   &0.002  \\
   $C_{m}$    &0.0084& $\pm$  &    0.0011&    0.0175& $\pm$   &0.0015  \\
 \hline
\end{tabular}

  \caption{$ 2\, \sigma_{od}$ for $q=6 \; {\rm and} \; 7$}
  \label{table1}

\end{center}
\end{table}

%$\;$

These scaling studies enable us to calculate the order-disorder
surface tension $2\sigma_{od}$.  The surface tensions $2\sigma_{od}$
of $q=6$ and $q=7$ state Potts models are calculated from the
autocorrelation time of the relaxation processes for the
observables. In Table 1. we have presented $2\,\sigma_{od}$ which are
obtained from the relaxation of three different quantities. Depending
on the quantity, the value of the surface tension is observed to vary
slightly. Nevertheless, the surface tension, within errorbars, is
$0.008\pm0.001$ and $0.017\pm0.004$ for $q=6$ and $q=7$ respectively.
The error on the surface tension can be taken as the fluctuation of 
the values obtained using different operators.

\section{Conclusions}

In conclusion we have numerically simulated the dynamic relaxation
process of the two-dimensional $q=6$ and $7$ state Potts models
starting from random initial states with vanishing initial order
parameter.  Here in this preliminary work we have investigated the
dynamical scaling properties of the first-order phase
transitions. This work is based on two well established facts that the
autocorrelation time of the critical relaxations in first-order phase
transitions are given by the instanton calculations~\cite{Niel:1987}
and all infinities of the thermodynamic quantities are governed by the
volume of the system~\cite{Borgs:1990,Borgs:1991,Billoire:1995}.
Under these assumptions one may expect that any thermodynamical
quantity exhibits dynamical scaling considering the correction to
scaling terms.

$\;$

We  have demonstrated that for first-order phase transitions  a
universal scaling behaviour emerges already in the macroscopic
short-time regime  of the dynamical evolution. This scaling
behaviour resembles closely dynamic scaling which seems to exist
in second-order phase transitions. Furthermore, such a scaling
opens new and alternative methods of calculating surface tension
and it can be used to distinguish weak-first-order phase
transitions from the second-order one.

\pagebreak

$\;$

\pagebreak

\section*{Figure captions}

\begin{description}
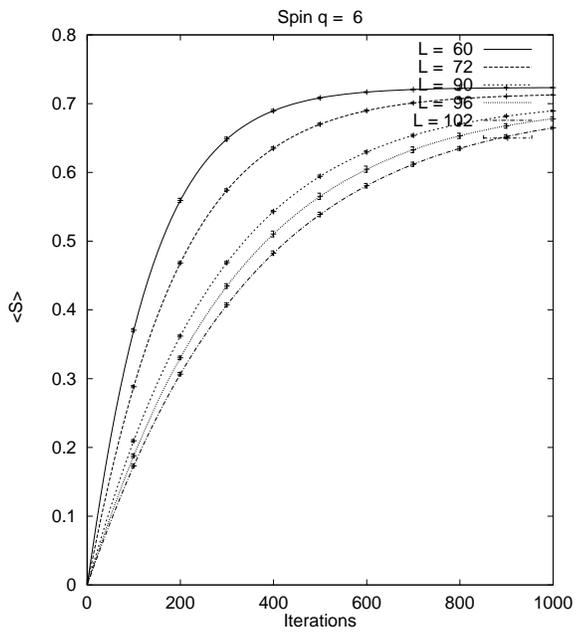
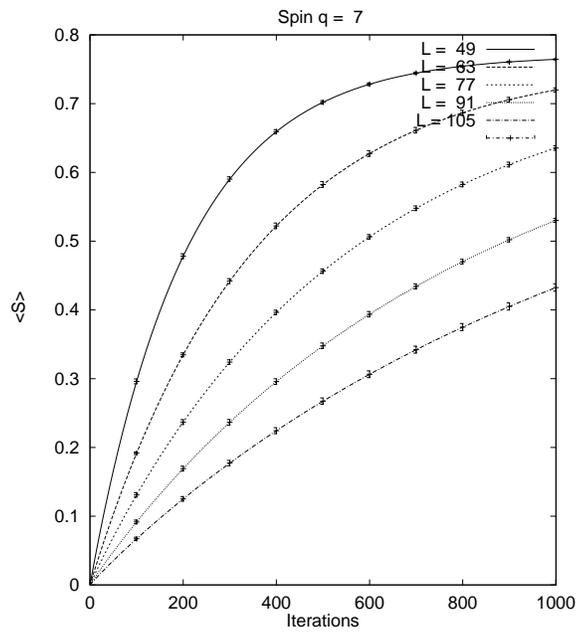
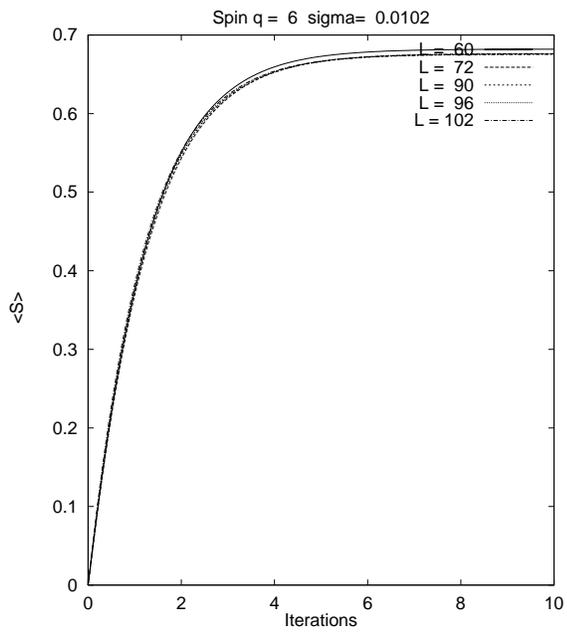
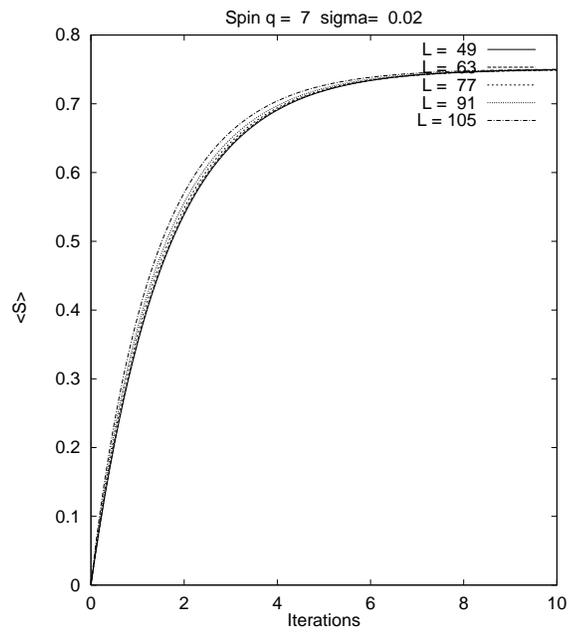
\item {Figure 1.}  (a) and (b) are the time evolution of the order 
                   parameter M, and (c) and (d) are their scaled form
                   for $q=6 \;{\rm and}\; 7$-state Potts model respectively. 
                   (The errorbars are omitted from the scaled forms 
                    for clarity of the figures.)

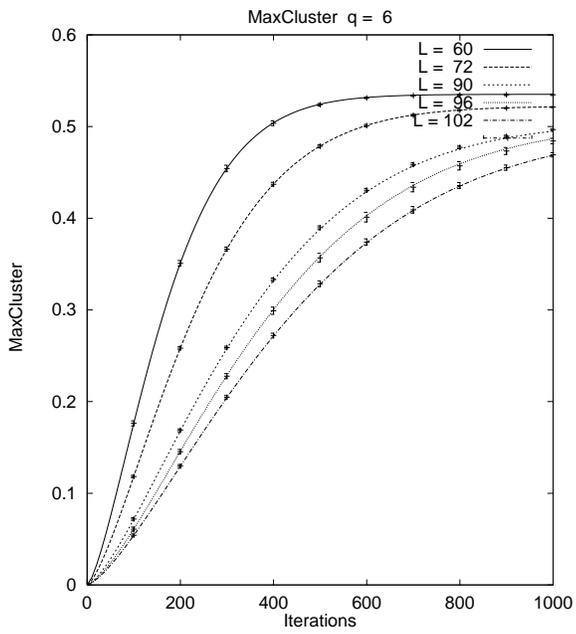
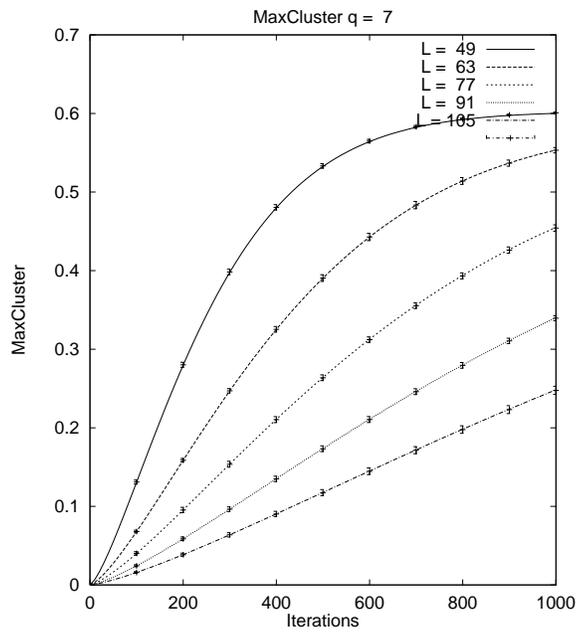
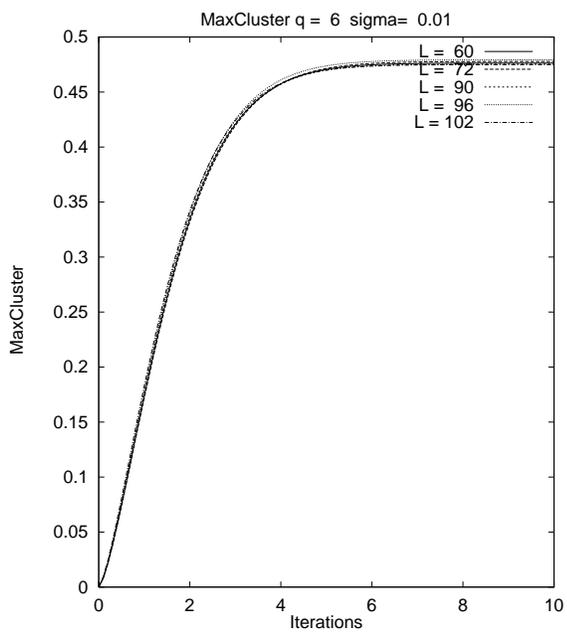
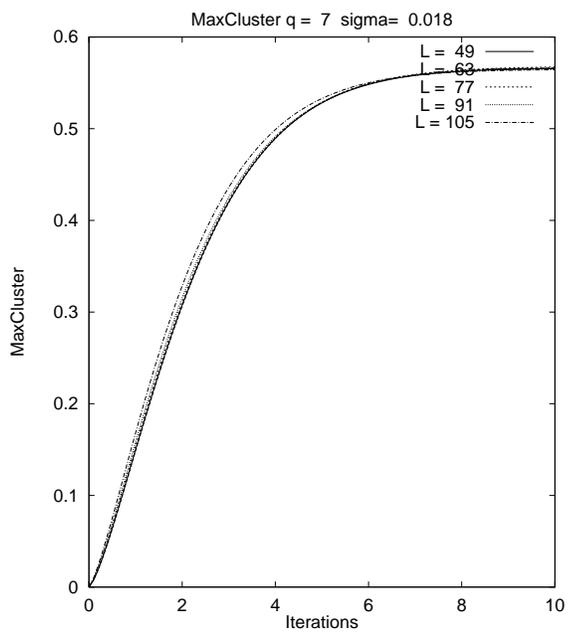
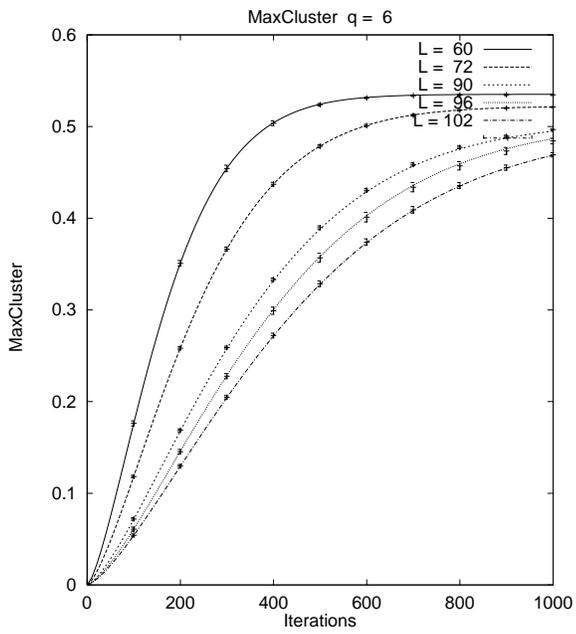
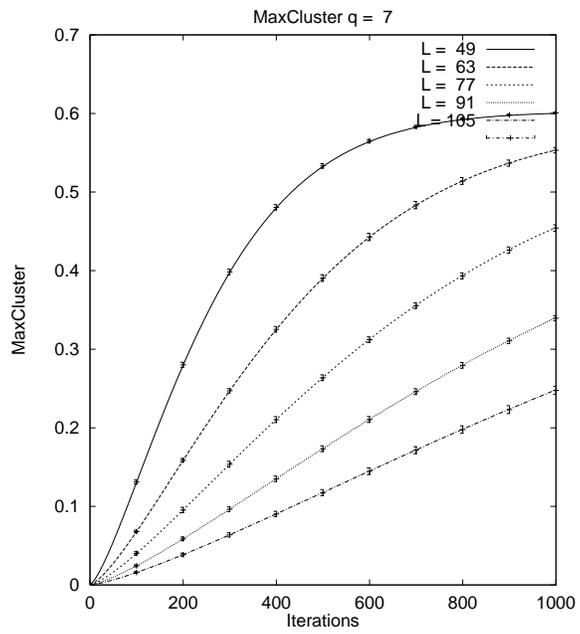
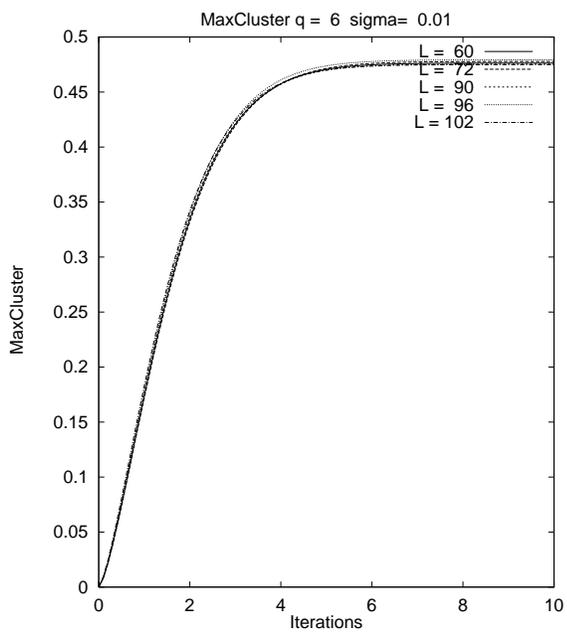
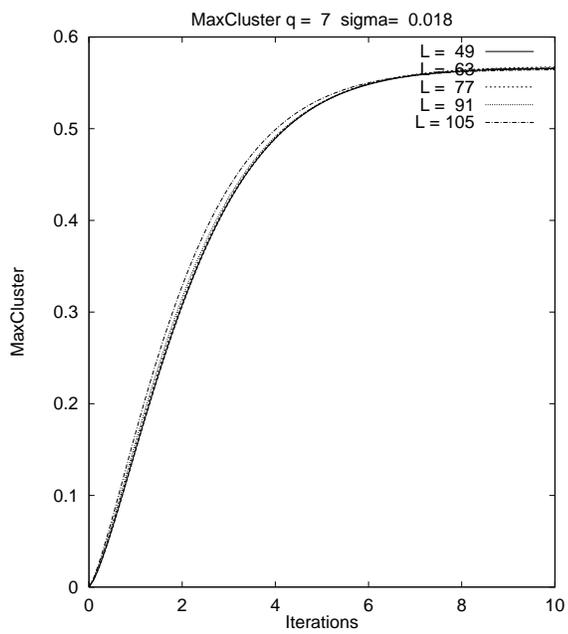
\item {Figure 2.} Same as fig. 1 but plots are for the maximum clusters.

\end{description}

\pagebreak


\begin{thebibliography}{99}

\bibitem{Janssen:1989}
       H.K.Janssen, B. Schoub and B. Schmittmann 
       Z. Phys. B73 (1989) 539.

\bibitem{Hohenberg:1977}
       P.C.Hohenberg and B. I. Halperin 
       Rev. Mod. Phys. B49 (1977) 435.

\bibitem{Zheng:1998}
       B. Zheng,
       Int. J. Mod. Phys. B 12 (1998) 1419.

\bibitem{Borgs:1990} 
       C. Borgs and R. Kotecky, 
       J. Stat. Phys. 61 (1990) 79.

\bibitem{Borgs:1991} 
       C. Borgs, R. Kotecky and S. Miracle-Sole,
       J. Stat. Phys. 62 (1991) 529.

\bibitem{Binder:1982}
       K. Binder,
       Phys. Rev. B 25 (1982) 1699.

\bibitem{Niel:1987}
       J. C. Niel, J. Zinn-Justin,
       Nucl. Phys. B280 (1987) 355.


\bibitem{Wang:1987} 
       R. H. Swendsen and J. S. Wang,
       Phys. Rev. Lett. 58 (1987) 86.

\bibitem{Wolff:1989} 
       U. Wolff,
       Phys. Rev. Lett. 62 (1989) 361.

\bibitem{Aydin:1996}
       M. Ayd{\i}n and Y. G\"{u}nd\"{u}\c{c}
       Physica A 232 (1996) 265.

\bibitem{Yasar:1998}
       F. Yasar, Y. G\"{u}nd\"{u}\c{c}, M. Ayd{\i}n and T. \c{C}elik
       Physica A 255 (1998) 430.

\bibitem{Binder:1981}
       K. Binder,
       Phys. Rev. Lett. 47 (1981) 693.

\bibitem{Potts:1952} 
       R. B. Potts, 
       Proc. Camb. Phil. Soc. 48 (1952) 106.

\bibitem{Wu:1982} 
       F. Y. Wu, 
       Rev. Mod. Phys. 54 (1982) 235.

\bibitem{Baxter:1973} 
       R. J. Baxter, 
       J. Phys. C6 (1973) L445.

\bibitem{Buffenoir:1993} 
       E. Buffenoir and S. Wallon, 
       J. Phys. A26 (1993) 3045.

\bibitem{Billoire:1995} 
       A. Billoire, 
       Nucl. Phys. (Proc. Suppl.) B42 (1995) 21.

\end{thebibliography}
\end{document}